\documentclass[11pt,draftcls,onecolumn] {IEEEtran}

\usepackage{amssymb,amsmath,amsfonts}
\usepackage{mathrsfs}
\usepackage{array} 

\newtheorem{theorem}{Theorem}

\newtheorem{corollary}{Corollary}

\newcommand{\supp}[1]{\texttt{supp}(#1)}
\newcommand{\suppk}[1]{\texttt{supp}_{k}(#1)}

\newcommand{\pedec}{\text{P}_\text{err}(\mathcal{D})}
\newcommand{\pe}{\text{P}_\text{err}}

\newcommand{\expected}[1]{\mathbb{E}_{#1}}
\newcommand{\mineig}{\sigma_\text{min}}
\newcommand{\mean}[1]{\text{E}\{#1\}}

\date{20 May 2013}

\begin{document}
\title{Tight Sufficient Conditions on Exact Sparsity Pattern Recovery}

\author{Behrooz~Kamary~Aliabadi and Sil\`{e}ye Ba
\thanks{Behrooz~Kamary~Aliabadi is with the Electronics department of T\'{e}l\'{e}com Bretagne (Institut Mines-T\'{e}l\'{e}com),
CS 83818, 29238 Brest, France.
He is also with Universit\'e Europ\'eenne de Bretagne (UEB)
and the Laboratory for Science and Technologies of Information, Communication and Knowledge, UMR CNRS 6285 Lab-STICC, Brest, France
(e-mail: behrooz.kamaryaliabadi@telecom-bretagne.eu). Sil\`{e}ye Ba is with RN3DLab Innovation Lab (sileye.ba@rn3dlab.com).}}
\maketitle
\begin{abstract}
A noisy underdetermined system of linear equations is considered
in which a sparse vector (a vector with a few nonzero elements)
is subject to measurement. The measurement matrix elements
are drawn from a Gaussian distribution.
We study the information-theoretic constraints on exact support
recovery of a sparse vector from the measurement vector and matrix.
We compute a tight, sufficient condition that is applied
to ergodic wide-sense stationary sparse vectors.
We compare our results with the existing bounds and recovery
conditions. Finally, we extend our results to approximately sparse signals.
\end{abstract}


\section{Introduction}
\label{sec:introduction}
Solving underdetermined systems of linear equations appears in various applications.
In general, they have an infinite number of solutions. Recent studies in
\cite{Candes2008b,Donoho2006} show that
these systems of linear equations have unique solutions
(if the solution is sparse and the system is noise-free)
under certain conditions.

We consider a noisy system of linear equations for which it is \emph{a priori} known
that the solution is $k$-sparse (a vector with $k$ nonzero elements).
\begin{equation}
Y = X\beta + W
\label{eq:system}
\end{equation}
where $X \in \mathbb{R}^{n \times p}$ is a random Gaussian measurement matrix
with independently and identically distributed (i.i.d.) elements $X_{ij} \sim \mathcal{N}(0,1)$.
$\beta$ is the $k$-sparse vector subject to measurement.
$W$ is a Gaussian noise vector $W \sim \mathcal{N}(0,I_{n\times n})$.
We define the support set of $\beta$ as 
\[
\supp{\beta} \triangleq \{i : \beta_i \neq 0\},
\]
that is a set of indices where elements of $\beta$ are nonzero.
The estimation of $\beta$ as a function of $X$ and $Y$ is an inverse problem that consists
of (a) detecting the support and (b) estimating the amplitudes of the nonzero elements \cite{Tarokh2010,Rad2011}.
Once the support set of $\hat{\beta}$ is determined, the estimated sparse (optimal) solution is
\begin{equation}
\hat{\beta} = \underset{\nu}{\operatorname{arg\ min}}\, ||Y - X_{\supp{\hat{\beta}}}\nu||_2^2
\label{eq:mse-decoder}
\end{equation}
where $X_{\supp{\hat{\beta}}}$ is a $n \times k$ sub-matrix of
the measurement matrix with column indices in $\supp{\hat{\beta}}$.
Therefore, as it is discussed in \cite{Tarokh2010,Rad2011,Wang2010}, finding
the optimal solution of such a noisy system is reduced to the exact support recovery.

The error metric is the 0\---1 loss function defined in \cite{Wainwright2009,Wang2010,Tarokh2010}
\begin{equation}
\begin{split}
\rho(\beta,\hat{\beta}) =\ & \mathbb{I} \left [\left \{ \hat{\beta}_i \neq 0, \, \forall i \in \supp{\beta} \right \}\right .\\
& \left . \cap \left \{ \hat{\beta}_j = 0, \, \forall j \not \in \supp{\beta} \right \} \right ]
\end{split}
\end{equation}
where $\mathbb{I}(.)$ is the indicator function.
We define a decoder $\mathcal{D}:Y\rightarrow \theta$ that maps the vector $Y$
to a support set $\theta = \supp{\hat{\beta}}$.
The probability of choosing a wrong support set is $\text{Pr}[\theta \neq \supp{\beta} | X, \beta]$
over the measurement noise and the sampling matrix.
The average detection error is defined as in \cite{Wainwright2009,Wang2010}
$$\pe = \frac{1}{\binom{p}{k}} \text{Pr} [\mathcal{D}(Y) \neq \supp{\beta} | X, \beta]$$
where the support $\supp{\beta}$ is assumed to be chosen uniformly random from $\binom{p}{k}$
possible subsets of size $k$ \cite{Wainwright2009,Wang2010}. In an exact support recovery regime,
we intend to have asymptotically zero average detection error where
$\pe \rightarrow 0$ as $n \rightarrow \infty$ \cite{Wainwright2009,Wang2010}.

In this paper, we compute asymptotic sufficient conditions (on exact support recovery) depending on the number of measurements $n$,
the sparse vector dimension $p$, the sparsity level $k$ and the signal-to-noise ratio 
\begin{equation}
\mathrm{SNR} = \frac{\mean{||X\beta||_2^2}}
{\mean{||W||_2^2}} = ||\beta||_2^2,
\end{equation}
where the noise variance $\sigma^2 = 1$ and the measurement matrix elements
are drawn from a Gaussian random source output with unit variance and zero mean.
Though the signal-to-noise ratio is an important parameter, the exact support recovery of a $k$-sparse vector is not solely
guaranteed by its SNR \cite{Wang2010,Rad2011}. We consider the case in which the decoder has the
highest failure probability for a given SNR, by taking $|\beta_i| = \lambda$ where $i \in \supp{\beta}$ and
$\lambda$ is the minimum absolute value of any nonzero element of the strictly sparse vector $\beta$.
By such assumption, the recovery of any $k$-sparse vector with $\text{SNR} \geq k\lambda^2$ is
guaranteed \cite{Wainwright2009,Wang2010,Rad2011}.

In \cite{Wang2010} the authors assume that $\beta$ has a mean and variance stationary source.
In \cite{Baron2010,Shihao2008} the authors model
the support of a sparse signal as a vector of random elements, in which an element is an outcome
of a random source with probability $k/p$ to be nonzero that implies mean stationarity.
In \cite{Baron2011,Jalali2012} the authors model the high-dimensional sparse vector $\beta$
as a realization of an ergodic stationary source.
In signal processing, random sources are widely modeled as ergodic wide-sense stationary
(EWSS) \cite{Hayes1996,Haykin2001}. In this work the sparse vector $\beta$ is assumed to be random
ergodic wide-sense stationary. Even though this assumption is common and most often inevitable in signal processing, it
has not been considered in \cite{Wainwright2009,Wang2010,Rad2011,Aeron2010,Tarokh2010,Sarvotham2006,Jin2011,Fletcher2009}
to compute the information-theoretic constraints.

The rest of this article is organized as follows. The main results are given in section~\ref{sec:results}.
In section~\ref{sec:strictly-sparse-signals} we give sufficient conditions for EWSS strictly
sparse vectors and compare it with the existing necessary and sufficient results. By the comparison
we observe that the derived constraints are tighter than the existing bounds and conditions (where the sparse vector is EWSS).
In section~\ref{sec:approximately-sparse-signals} the given results are extended to
approximately sparse and ergodic wide-sense stationary signals. In section~\ref{sec:proof} the proof of the theorems
and corollaries are given. Finally, in section \ref{sec:conclusion} the conclusion is given.

\section{Results}
\label{sec:results}

In \cite{Wainwright2009,Wang2010,Aeron2010} and this article, Fano's inequality \cite{Cover2006} is exploited
to obtain asymptotic constraints on the exact support recovery depending on $(p,k,n,\lambda)$.
Fano's inequality in asymptotic form given in \cite{Wang2010,Rad2011} is
\begin{equation}
\label{eq:fano-asymp}
\log \left ( \binom{p - k + m}{m} - 1 \right ) - \log 2 \leq \expected{X} I(\theta;Y),
\end{equation}
for $m = 1, \dots,k$ where $k$ is the sparsity level, $p$ is the sparse vector dimension and $\expected{X} I(\theta;Y)$
is the expected information rate between measurement vector $Y$ and detected support vector $\theta$ (see \cite{Aeron2010,Wang2010}).
$m$ is the number of nonzero elements that is not known by the support detector, i.e., the detector
\emph{a priori} knows indices of $k - m$ nonzero (or significantly large) elements.
As it is discussed in \cite{Wang2010} having \emph{a priori} knowledge about some nonzero elements
may not facilitate the support detection. In fact, for some $(p,k,m)$ where $m < k$ there are more support
sets to choose than $m=k$. The cardinality of the support sets
that may be chosen wrongly by the support detector is $\binom{p - k + m}{m} - 1$.
This appears in the left-hand side of \eqref{eq:fano-asymp} that reaches a maximum value for a particular $(p,k,m)$
due to its concavity (see the discussion in \cite{Wang2010}).

\subsection{Strictly Sparse Signals}
\label{sec:strictly-sparse-signals}
The recovery constraints in \cite{Wainwright2009,Wang2010,Rad2011,Aeron2010,Tarokh2010,Sarvotham2006,Jin2011,Fletcher2009}
are dedicated to strictly sparse signals. These signals have explicitly $k$ nonzero elements and the rest are zeros.

The tightness of the recovery constraints through~\eqref{eq:fano-asymp} depends on its right-hand side
that is $\expected{X} I(\theta;Y)$ (see \cite{Wang2010,Aeron2010}).
To compute the information rate, the autocorrelation matrix of the sparse vector is required (see \cite{Wang2010,Aeron2010,Sarvotham2006}).
In general, the autocorrelation matrix is computed from the probability distribution function (PDF) of the process.
The distribution function may not always be known and assuming a specific PDF restricts the given constraint
to that specific random process.
Therefore, in \cite{Sarvotham2006,Aeron2010} an upper bound on $I(\theta;Y)$ is computed and exploited to obtain their recovery constraints.
In \cite{Wang2010} the authors compute a tighter upper bound (with respect to \cite{Sarvotham2006,Aeron2010}) on the information rate that
results in tighter necessary conditions on the exact support recovery.

Where the process is ergodic and wide-sense stationary (EWSS), the autocorrelation function and consequently
the autocorrelation matrix can be computed from its time realizations \cite{Papoulis1991,Hayes1996}.
The random sparse vector is chosen uniformly from ${p \choose k}$ possible $k$-sparse vectors of dimension $p$
in which a vector element is nonzero with probability $k/p$ \cite{Papoulis1991,Baron2011,Shihao2008}.
This is employed in \cite{Baron2011,Shihao2008} to compute different statistics and sparse signals.
EWSS assumption is exploited to compute the autocorrelation matrix for the worst case scenario
(where $|\beta_i| = \lambda\; \forall i \in \supp{\beta}$) without having knowledge of PDF.
Having the autocorrelation function, the exact information rate can be computed for the worst case scenario EWSS signals.
What remains as an obstacle is the computation of the expected information rate in~\eqref{eq:fano-asymp}
with respect to sampling matrix $X$ (for which only its distribution is known). To overcome that,
the combination of Jensen and Minkowski inequalities is exploited to obtain a lower bound on $\expected{X} I(\theta;Y)$.
It is worth mentioning that the combination of Jensen and Minkowski inequalities for the information rate results in extremely tight bounds
that is reported in \cite{Oyman2002,Jin2005}. This combination is of great importance since
it let us compute the information rate through product of two polytopes' volumes. The
first is a Wishart matrix (that defines a polytope) constructed from the Gaussian
sampling matrix and the second is the autocorrelation matrix of the worst case ergodic wide-sense stationary signal.
This tight lower bound on the expected information rate is used in~\eqref{eq:fano-asymp} to get a sufficient condition
depending on $(p,n,k,\lambda)$ which is given in the following theorem.
\begin{theorem}
\label{thm:sufficient-condition-tight}
Assume a measurement matrix $X \in \mathbb{R}^{n\times p}$ whose elements are drawn from the outcome 
of an i.i.d. Gaussian random source with zero mean and unit variance, i.e.,
$X_{i,j} \sim \mathcal{N}(0,1)$. A sufficient condition for asymptotically
reliable recovery of a $k$-sparse ergodic wide-sense stationary signal $\beta$
in which the nonzero elements $|\beta_i| \geq \lambda$ is
\begin{equation}
\log \left[{p - k + m \choose m} - 1\right] - 1 \leq L
\label{eq:fano-inequality}
\end{equation}
where
\begin{equation*}
\begin{split}
L = \frac{n}{2}\log\left [ 1 + \frac{m}{p - k + m} \lambda^2 (1 - \frac{m}{p - k + m}) \right .\\
\left . \sqrt[n]{\Gamma(n)\,{p - k + m - 1 \choose n - 1}} \right ]
\end{split}
\end{equation*}
in which $\Gamma(.)$ is Gamma function and $m = 1, \dots, k$.
\end{theorem}

By lower bounding further the information rate (right-hand side of \eqref{eq:fano-inequality}), the following sufficient condition is obtained.
\begin{corollary}
\label{cor:sufficient-condition}
Assume a measurement matrix $X \in \mathbb{R}^{n\times p}$ whose elements are drawn from the outcome 
of an i.i.d. Gaussian random source with zero mean and unit variance, i.e.,
$X_{i,j} \sim \mathcal{N}(0,1)$. A sufficient condition for asymptotically
reliable recovery of a $k$-sparse ergodic wide-sense stationary signal $\beta$
in which the nonzero elements $|\beta_i| \geq \lambda$ is
\begin{equation}
n > \max \{ f_1(p,k,\lambda), \dots, f_k(p,k,\lambda),k\}
\end{equation}
where
\begin{equation}
f_m(p,k,\lambda) = \frac{\log \left[ {p - k + m \choose m} - 1\right] - 1}{\frac{1}{2}\log \left(1 + \frac{m}{e}\lambda^2 \left ( 1 - \frac{m}{p - k + m}\right)\right)}
\end{equation}
for $m = 1, \dots, k$.
\end{corollary}

For the sake of completeness and comparison we rewrite the necessary condition given by Wang~\emph{et al.}
\begin{theorem}{\cite{Wang2010}}
\label{thm:wang}
Assume the measurement matrix $X \in \mathbb{R}^{n\times p}$ whose elements are drawn
from the outcome of an i.i.d. Gaussian source with zero mean and unit variance.
A necessary condition for asymptotically reliable recovery of a $k$-sparse signal $\beta$ in which
nonzero elements $|\beta_i| \geq \lambda$ is
\begin{equation}
n > \max \{ f_1(p,k,\lambda), \dots, f_k(p,k,\lambda),k\}
\end{equation}
where
\begin{equation}
f_m(p,k,\lambda) = \frac{\log {p - k + m \choose m} - 1}{\frac{1}{2}\log \left(1 + m\lambda^2 \left ( 1 - \frac{m}{p - k + m}\right)\right)}
\end{equation}
for $m = 1, \dots, k$.
\end{theorem}

In Table~\ref{tbl:asymptotic} the given result for strictly sparse signals is compared with the sufficient condition in \cite{Rad2011}
and the necessary condition in \cite{Wang2010}. In this table the first three rows represent the constraints for linear sparsity regimes.
The last three rows of the table show the constraints for sublinear regimes. In the last two rows the signal-to-noise ratio tends to infinity
, i.e, $k\lambda^2\rightarrow \infty$.

Corollary~\ref{cor:sufficient-condition} is obtained by loosening further the information rate lower bound with respect to
right-hand side of \eqref{eq:fano-inequality} in Theorem~\ref{thm:sufficient-condition-tight}. In spite of this
we observe in Table~\ref{tbl:asymptotic} that Corollary~\ref{cor:sufficient-condition} is as tight as previous sufficient and necessary results.
From this we conclude that Theorem~\ref{thm:sufficient-condition-tight} may be asymptotically tighter than the results in \cite{Wang2010,Rad2011}.

The given tight sufficient condition in Corollary~\ref{cor:sufficient-condition} is valid for the whole
range of signal-to-noise ratio whereas the sufficient condition for exact support recovery in \cite{Rad2011}
is claimed to be restricted to $\lambda^2 = \Omega(\frac{1}{k})$ or $\lambda^2 = O(1)$ \cite{Rad2011}.


\subsection{Approximately Sparse Signals}
\label{sec:approximately-sparse-signals}
The practical signals are not strictly sparse but they are
approximately sparse with a few significantly large elements and the rest are small but
nonzero. In this set-up the existing constraints based on the sparsity level
may not work.

A feature of Fano's inequality is relating the detection error between the
source and the estimation random vectors. The source random vector (here it is the approximately sparse vector)
does not necessarily have a discrete alphabet but the estimated vector must be drawn from a discrete and countable
alphabet \cite{Cover2006}. This plays an important role in computation of recovery conditions for approximately sparse signals.

We assume that the wide-sense stationary and approximately sparse random vector $\beta$ has $k$ significantly
large elements and the rest are small and the support detector recovers the $k$ largest elements.
To obtain recovery conditions for the approximately sparse signals we assume a decoder
$\mathcal{D} : Y \rightarrow \theta$ where $\theta$ is a detected support set with cardinality $k$.
The support set of an approximately sparse signal is defined to be the indices of the $k$ largest elements 
\[
\suppk{\beta} \triangleq \{i : \beta_i \neq 0, \beta_i > \beta_{j} > \beta_{p}, k+1\leq j\leq p \}.
\]
The error metric is defined as
\begin{equation}
\begin{split}
\rho_k(\beta,\hat{\beta}) =\ & \mathbb{I} \left [\left \{ \hat{\beta}_i \neq 0, \, \forall i \in \suppk{\beta} \right \}\right .\\
& \left . \cap \left \{ \hat{\beta}_j < \beta_{k}, \, \forall j \not \in \suppk{\beta} \right \} \right ]
\end{split}
\end{equation}
where $\beta_k$ is the $k$th largest element of $\beta$.
The probability of choosing a wrong support set is $\text{Pr}[\theta \neq \suppk{\beta} | X, \beta]$ and
the average detection error is 
$$\pe = \frac{1}{{p \choose k}} \text{Pr} [\mathcal{D}(Y) \neq \suppk{\beta} | X, \beta]$$
where $\suppk{\beta}$ is the support set that is uniformly chosen from ${p \choose k}$ subsets
of size $k$ \cite{Wainwright2009,Wang2010}.

In this problem a similar approach to strictly sparse signals is used. The combination of Jensen and Minkowski
inequalities are used to obtain a lower bound on the expected information rate in~\eqref{eq:fano-asymp}.
For the approximately sparse signals it may not be possible to compute the autocorrelation matrix from $(p,n,k)$
and $\lambda$ can not generally be assumed constant. A lower bound on the volume of the polytope defined
by the autocorrelation matrix can be alternatively computed from the signal power spectrum \cite{Hayes1996}.
Therefore, we assume that the power spectrum of the signal is in hand through measurement and estimation.
Having the power spectrum of the signal, we compute the information rate lower bound and obtain a sufficient condition as follows.
\begin{theorem}
\label{thm:approximately-sparse}
Assume a measurement matrix $X \in \mathbb{R}^{n\times p}$ whose elements are drawn from the outcome 
of an i.i.d. Gaussian random source with zero mean and unit variance, i.e., $X_{i,j} \sim \mathcal{N}(0,1)$.
A sufficient condition for asymptotically reliable recovery of a wide sense-stationary sparse signal $\beta$
with $k$ large nonzero elements is
\begin{equation}
\log \left[{p - k + m \choose m} - 1\right] - 1 \leq L
\label{eq:inequality}
\end{equation}
where
\begin{equation}
L = \frac{n}{2}\log\left [ 1 + G\;\sqrt[n]{\Gamma(n)\,{p - k + m - 1 \choose n - 1}} \right ]
\label{eq:approximately-sparse-l}
\end{equation}
in which $\Gamma(.)$ is Gamma function, $m = 1, \dots, k$ and
$G$ is the infimum of the approximately sparse signal power spectrum. 
\end{theorem}
\begin{corollary}
\label{cor:full}
Assume $k = \Theta(p)$. In Theorem~\ref{thm:approximately-sparse} the sufficient condition for asymptotically
reliable recovery of a wide-sense stationary and approximately sparse signal $\beta$ is obtained by replacing
\begin{equation}
G = \frac{1}{2\pi} \int_{0}^{2\pi} \log S(\omega)\; \mathrm{d}\omega
\end{equation}
\end{corollary}
in \eqref{eq:approximately-sparse-l} where $S(\omega)$ is the power spectrum of the sparse signal.

\begin{table*}[htp!]
\centering
\caption{Sufficient and Necessary Conditions on the Number of Measurements $n$ for Exact Support Recovery in the Linear and the Sublinear Regimes.}
\label{tbl:asymptotic}
\begin{tabular}{|l|c|c|c|}
\hline
Scaling & Sufficient Condition Corollary~\ref{cor:sufficient-condition} & Necessary Condition Theorem~\ref{thm:wang} \cite{Wang2010} & Sufficient Condition \cite{Rad2011}\\
\hline \hline
\parbox[t]{2cm}{$k = \Theta(p)$\\ $\lambda^2 = \Theta(\frac{1}{k})$} & $n = \Theta(p\log p )$ & $n = \Theta(p\log p)$ & $n = \Theta(p\log p)$\\
\hline
\parbox[t]{2cm}{$k = \Theta(p)$\\ $\lambda^2 = \Theta(\frac{\log k}{k})$} & $n =\Theta(p)$ & $n = \Theta(p)$ & $n = \Theta(p)$\\
\hline
\parbox[t]{2cm}{$k = \Theta(p)$\\ $\lambda^2 = \Theta(1)$} & $n = \Theta(p)$ & $n = \Theta(p)$ & $n = \Theta(p)$\\
\hline
\hline
\parbox[t]{2cm}{$k = o(p)$\\ $\lambda^2 = \Theta(\frac{1}{k})$} & $n = \Theta(k\log(p - k))$ & $n = \Theta(k\log(p - k))$& $n = \Theta(k\log(p - k))$\\
\hline
\parbox[t]{2cm}{$k = o(p)$\\ $\lambda^2 = \Theta(\frac{\log k }{k})$} &
$n = \max \left\{ \Theta(\frac{k\log(p-k)}{\log k}),\Theta(\frac{k\log \frac{p}{k}}{\log \log k}) \right\}$ &
$n = \max \left\{ \Theta(\frac{k\log(p-k)}{\log k}),\Theta(\frac{k\log \frac{p}{k}}{\log \log k}) \right\}$ &
$n = \max \left\{ \Theta(\frac{k\log(p-k)}{\log k}),\Theta(\frac{k\log \frac{p}{k}}{\log \log k}) \right\}$\\
\hline
\parbox[t]{2cm}{$k = o(p)$\\ $\lambda^2 = \Theta(1)$} &
$n = \max \left\{ \Theta(\frac{k\log \frac{p}{k}}{\log k}),\Theta(k) \right\}$ &
$n = \max \left\{ \Theta(\frac{k\log \frac{p}{k}}{\log k}),\Theta(k) \right\}$ &
$n = \max \left\{ \Theta(\frac{k\log \frac{p}{k}}{\log k}),\Theta(k) \right\}$\\
\hline
\end{tabular}
\end{table*}
\section{Proof}
\label{sec:proof}
\subsection{Theorem~\ref{thm:sufficient-condition-tight}}
To obtain the exact support recovery conditions on the number of measurements,
we exploit Fano's inequality
\begin{equation}
\pe \geq 1 - \frac{I(\theta;Y) + \log 2}{\log \left ({p \choose k} - 1 \right )}
\label{eq:fano}
\end{equation}
where $I(\theta;Y)$ is the mutual information between $\theta$ and $Y$
\cite{Wainwright2009,Wang2010,Aeron2010,Cover2006}. $\theta$ is the detected support
set that is given by the decoder,  $\theta = \mathcal{D}(Y)$. $\pe$ is the probability
that the decoder fails to detect the correct support set, i.e.,
$$\pe = \text{Pr}[\supp{\beta} \neq \theta|X,\beta].$$
The error probability of the decoder, $\pedec$, is the average error probability with respect to
Gaussian measurement matrix $X$.
This is $$\pedec = \expected{X} \pe$$
where $\expected{X}$ denotes the expected value operator with respect to $X$ 
\cite{Tarokh2010,Wang2010}.
We further assume that the decoder has \emph{a priori} knowledge of all
 nonzero locations of $\beta$ but $m$ locations with smallest values
($1 \leq m \leq k $). The decoder has to choose from ${p - k + m \choose m}$
support sets \cite{Wang2010}. Let $\mathcal{U}$ be the set of unknown location
indices where $|\mathcal{U}| = m$. The $n$-dimensional observation vector is
$$\tilde{Y} = \tilde{X}\tilde{\beta} + W$$
where $\tilde{X}$ is the measurement matrix with column indices in $\mathcal{U}$.
$\tilde{\beta}$ is the vector subject to measurement with element indices in $\mathcal{U}$. 
Therefore, the error probability of the decoder is bounded as
\begin{equation}
\pedec \geq 1 - \frac{ \expected{\tilde{X}} I(\theta;\tilde{Y}) + \log 2}{\log \left ({p - k + m\choose m} - 1 \right )}.
\label{eq:fano-expected}
\end{equation}
The mutual information $I(\theta;\tilde{Y})$ is given by
\begin{equation}
\label{eq:info-rate}
\begin{split}
 I(\theta;\tilde{Y}) &= H(\tilde{Y}|\tilde{X}) - H(\tilde{Y}|\theta,\tilde{X})\\
                         &= H(\tilde{Y}|\tilde{X}) - H(W)\\
                         &= \frac{1}{2} \log \left | I_n + \tilde{X} R_{\tilde{\beta}} \tilde{X}^\dag \right |
\end{split}
\end{equation}
where $R_{\tilde{\beta}}$ is the autocorrelation matrix of random vector $\tilde{\beta}$.
The equality $\left| I_p + AB \right| = \left| I_n + BA \right|$ holds for any pair of matrices $A_{p\times n}$
and $B_{n\times p}$ \cite{Meyer2001}. This is an algebraic equality that implies equality of the volumes.
This is required to separate the autocorrelation matrix from the sampling matrix and its conjugate transpose that faciliates
the computation of the information rate.
Therefore, \eqref{eq:info-rate} can be rewritten as
\begin{equation}
I(\theta;\tilde{Y}) = \frac{1}{2} \log \, \left| I_{p-k+m} + \tilde{X}^\dagger \tilde{X} R_{\tilde{\beta}} \right|.
\label{eq:info-rate-prime}
\end{equation}
We have $\text{rank}(\tilde{X}^\dagger \tilde{X} R_{\tilde{\beta}}) \leq \min \{ \text{rank}(\tilde{X}^\dagger \tilde{X}), \text{rank}(R_{\tilde{\beta}}) \}.$
In \cite{Feng2007} it is shown that $\text{rank}(X^\dagger X) = n$, which implies
$\min \{ \text{rank}(X^\dagger X), \text{rank}(R_\beta) \} = n$.

We get a lower bound on the mutual information by applying the Brunn-Minkowski
inequality \cite{Cover2006},
\begin{equation}
\begin{split}
I(\theta;\tilde{Y}) &\geq \frac{n}{2} \log \left ( 1 + | \tilde{X}^\dagger X R_{\tilde{\beta}} |^{1/n} \right )\\
                 &= \frac{n}{2} \log \left ( 1 + \exp \left ( \frac{1}{n} \log | \tilde{X}^\dagger \tilde{X} R_{\tilde{\beta}} |\right ) \right ).
\end{split}
\label{eq:cap-mdim}
\end{equation}
Now, by taking the expectation in \eqref{eq:cap-mdim} and by using Jensen's inequality we can write
\begin{equation}
\expected{\tilde{X}} I(\theta;\tilde{Y}) \geq \frac{n}{2} \log_2
\left [ 1 +   \exp \left ( \frac{1}{n}(a + b) \right ) \right ]
\label{eq:mean-cap-lb}
\end{equation}
where
\begin{equation}
a = \expected{\tilde{X}} \sum_{i=1}^{n} \log \sigma_i \left ( \tilde{X}^\dagger \tilde{X} \right )
\label{eq:a}
\end{equation}
and
\begin{equation}
b = \sum_{i=1}^{n} \log \sigma_i \left (R_{\tilde{\beta}} \right ).
\label{eq:b}
\end{equation}
$\sigma_i(.)$ is the $i$th eigenvalue of a matrix. In \eqref{eq:b}, the eigenvalues are sorted
$\sigma_1(R_{\tilde{\beta}}) < \sigma_2(R_{\tilde{\beta}}) < \dots < \sigma_{p - k + m}(R_{\tilde{\beta}})$.

\eqref{eq:a} is the expected value of
the logarithm of random Wishart matrix determinant $|\tilde{X}^\dagger \tilde{X}|$.
The product of the eigenvalues of this random Wishart matrix is distributed as product
of $n$ chi-square distributed random variables \cite{Lapidoth2003,Oyman2002},
and therefore we have
\begin{equation}
\prod_{i=1}^{n} \sigma_i \left ( \tilde{X}^\dagger \tilde{X} \right ) \sim \prod_{j=1}^{n}
\chi_{p - k + m - j + 1}^2.
\label{eq:chisquare}
\end{equation}
By taking the logarithm and then the expectation of the right-hand side of \eqref{eq:chisquare}
we obtain
\begin{equation}
a = \expected{} \sum_{j=1}^{n} \log \chi_{p - k + m - j + 1}^2 = -n\gamma + \sum_{j=1}^{n} \sum_{\ell=1}^{p - k + m - j} \frac{1}{\ell} 
\label{eq:sum-chi-square}
\end{equation}
where  $\gamma$ is  Euler's constant \cite{Lapidoth2003,Oyman2002}.
In the asymptotic regime where $\pe(\mathcal{D}) \rightarrow 0$ as $n \rightarrow \infty$ that implies $p \rightarrow \infty$ we have
\[
\gamma = \lim_{p\rightarrow \infty} \left [ \displaystyle \sum_{\ell = 1}^{p - k + m - j} \frac{1}{\ell} - \log (p - k + m - j) \right ].
\]
We then compute
\begin{equation}
\label{eq:log-det-gaussian}
\begin{split}
a &= \displaystyle \sum_{j=1}^{n} \log(p - k + m - j)\\
  &= \log(p - k + m - j - 1)! - \log(p - k + m - n - 1)!\\
  &= \log \frac{\Gamma(p - k + m)}{\Gamma(p - k + m - n)}\\
  &= \log \frac{\Gamma(p - k + m)}{\Gamma(n) \Gamma(p - k + m - n)} \Gamma(n)\\
  &= \log {p - k + m  - 1 \choose n - 1} + \log \Gamma(n)
\end{split}
\end{equation}
where $\Gamma(.)$ is Gamma function. We model the sparse vector elements $\tilde{\beta}_i$ as output of
an ergodic wide-sense stationary random vector source. Therefore, the elements of such random vector
is also ergodic wide-sense stationary. The nonzero elements appear with probability
$\text{Pr}[\tilde{\beta_i} \neq 0] = \frac{m}{p - k + m}$
\cite{Papoulis1991,Baron2010,Shihao2008}.
The autocorrelation matrix of $\tilde{\beta}$ is a Hermitian Toeplitz matrix
that is obtained from its autocorrelation function \cite{Hayes1996}.
We assume the nonzero elements $\tilde{\beta}_i$ where $i \in \mathcal{U}$,
are negative with probability $\xi$, i.e., $\xi = \text{Pr}[\tilde{\beta_i} \in \mathbb{R}^{-}]$.
The autocorrelation function at lag $\tau = 0$ is
\[
r_{\tilde{\beta}}(0) = \displaystyle \frac{1}{p - k + m} \sum_i \tilde{\beta}_i \tilde{\beta}_i = \frac{m}{p - k + m} \lambda^2.
\]
For lag $\tau \neq 0$ and $|\tau| \leq p$, the autocorrelation function is
\[
\begin{split}
r_{\tilde{\beta}}(\tau) &= \displaystyle \frac{1}{p - k + m} \sum_i \tilde{\beta}_i \tilde{\beta}_{i + \tau}\\
                             &= \frac{p - k + m - |\tau|}{p - k + m} (\frac{m}{p - k + m})^2\lambda^2\\
                             &\times \left [ \xi^2 + (1 - \xi)^2  - 2\xi(1-\xi)\right].
\end{split}
\]
For $|\tau| > p$ the autocorrelation function $r_{\tilde{\beta}}(\tau) = 0$. Therefore,
the autocorrelation function is
\[
r_{\tilde{\beta}}(\tau) = 
\begin{cases}
\frac{(p - k + m - |\tau|)m^2}{(p - k + m)^3} (4\xi^2 - 4\xi + 1)\lambda^2 & \text{if } |\tau| \leq p, \tau \neq 0\\
\frac{m}{p - k + m} \lambda^2 & \text{if } \tau = 0,\\
0 & \text{Otherwise}.
\end{cases}
\]
The autocorrelation matrix of $\tilde{\beta}$ is a Hermitian Toeplitz matrix with
$r_{\tilde{\beta}}(\tau)$ in its
first row, i.e., $$R_{\tilde{\beta}} = \texttt{Toeplitz}\{r_{\tilde{\beta}}\}.$$
We use $R_{\tilde{\beta}}$ and $r_{\tilde{\beta}}$ to find a lower bound on
\eqref{eq:b},
$$n \log \mineig \leq \sum_{i=1}^{n} \log \sigma_i \left (R_{\tilde{\beta}} \right )$$
where $\mineig \leq \sigma_i(R_{\tilde{\beta}})$ is the lower bound on all eigenvalues
of $R_{\tilde{\beta}}$. The minimum eigenvalue of $R_{\tilde{\beta}}$ is lower bounded
by the infimum of the power spectrum \cite{Hayes1996},
\begin{equation}
\mineig = \min_{\omega} S_{\tilde{\beta}}(\omega)
\label{eq:eiglb}
\end{equation}
where
\begin{equation}
\begin{split}
S_{\tilde{\beta}}(\omega) &= \left [ \frac{\sin (\omega (p + 1/2))}{\sin(\omega/2)}\right ]^2 (\frac{m}{p - k + m} \lambda)^2 (4\xi^2 - 4\xi + 1)\\
&+ \frac{m}{p - k + m}\lambda^2 - (\frac{m}{p - k + m}\lambda)^2 (4\xi^2 - 4\xi + 1)
\end{split}
\label{eq:ps}
\end{equation}
is the power spectrum of $\tilde{\beta}$. It is computed by taking Fourier transform of
$r_{\tilde{\beta}}$ \cite{Hayes1996}.
Therefore, we obtain
\begin{equation}
\mineig(\xi) = \left [ \frac{m}{p - k + m} - (\frac{m}{p - k + m})^2 (4\xi^2 - 4\xi + 1) \right ] \lambda^2
\end{equation}
from \eqref{eq:ps} and \eqref{eq:eiglb}. $\mineig$ is a function of $\xi$. The term
$0 \leq 4\xi^2 - 4\xi + 1 \leq 1$ reaches its maximum when $\xi \in \{0,1\}$, i.e., the sparse
vector subject to measurement is unipolar. To lower bound $\mineig$ we choose $\xi \in \{0,1\}$ that results in
\begin{equation}
\label{eq:log-mineig-lb}
b \geq n \log \left [(1 -  \frac{m}{p - k + m})\frac{m}{p - k + m}\lambda^2\right ].
\end{equation}

By substituting left-hand side of \eqref{eq:log-mineig-lb} and \eqref{eq:log-det-gaussian} in \eqref{eq:mean-cap-lb} we obtain
\begin{equation}
\begin{split}
\expected{X} I(\theta;\tilde{Y}) \geq \frac{n}{2}\log\left [ 1 + \sqrt[n]{\Gamma(n){p - k + m - 1 \choose n - 1}}\right .\\
\left . (1 -  \frac{m}{p - k + m})\frac{m}{p - k + m}\lambda^2\right].
\end{split}
\end{equation}
Finally, we obtain the sufficient condition as
\begin{equation}
\log \left[ {p - k + m \choose m}  - 1\right] - \log(2) \leq  L
\label{eq:lb}
\end{equation}
where
\begin{equation}
\begin{split}
L = \frac{n}{2}\log\left [ 1 + \sqrt[n]{\Gamma(n){p - k + m - 1 \choose n - 1}}\right .\\
\left . (1 -  \frac{m}{p - k + m})\frac{m}{p - k + m}\lambda^2\right]
\end{split}
\label{eq:rhs}
\end{equation}
for $m = 1, \dots, k$. 
\subsection{Corollary~\ref{cor:sufficient-condition}}
In proof of Theorem~\ref{thm:sufficient-condition-tight} we have
\begin{equation}
\label{eq:log-det-gaussian-cor}
a = \log {p - k + m - 1\choose n - 1} + \log \Gamma(n) 
\end{equation}
that can be lower bounded to obtain another lower bound on $\expected{X}I(\theta,\tilde{Y})$.
On the other hand we have
\begin{equation}
\sqrt{2\pi (n - 1) } (n - 1)^{(n - 1)} e^{-(n - 1)} \leq \Gamma(n)
\end{equation}
and
\begin{equation}
(n - 1) \log \left( \frac{p - k + m - 1}{n - 1} \right) \leq \log {p - k + m - 1\choose n - 1}.
\end{equation}
Considering $n \rightarrow \infty$ (see the asymptotic Fano's inequality in \cite{Wang2010}) and consequently $p \rightarrow \infty$
\begin{equation}
\label{eq:a-lb}
\begin{split}
a/n &\geq \frac{n - 1}{n} \log \frac{p - k + m - 1}{n - 1} \\
    &+ \frac{1}{n}\log \left[ \sqrt{2\pi(n - 1)} (n - 1)^{(n - 1)} e^{-(n - 1)}\right]\\
  &\geq \frac{n - 1}{n}\log(p - k + m - 1) - \frac{n - 1}{n}.
\end{split}
\end{equation}
By replacing the right-hand sides of \eqref{eq:a-lb} and \eqref{eq:log-mineig-lb} in \eqref{eq:mean-cap-lb} we have
\begin{equation}
L = \frac{n}{2} \log \left[1 + \frac{m}{e}(1 - \frac{m}{p - k + m})\lambda^2 \right]
\end{equation}
in \eqref{eq:lb}.
\subsection{Theorem~\ref{thm:approximately-sparse}}
In proof of Theorem~\ref{thm:sufficient-condition-tight} we assumed $\beta$ is strictly sparse. Where the signal is not strictly sparse we may not be able to
compute analytically a lower bound on \eqref{eq:b} as we computed in Theorem~\ref{thm:sufficient-condition-tight}.
In the proof of Theorem~\ref{thm:sufficient-condition-tight} we obtain a lower bound on $b$ through computation of a lower
bound on the minimum eigenvalue of the signal autocorrelation matrix.
Assume that we have the power spectrum of the approximately sparse signal in hand. The minimum eigenvalue of the autocorrelation matrix of the signal
is lower bounded by the infimum of the power spectrum (see \cite{Hayes1996} for details). Therefore, by replacing $\sigma_i(R_{\tilde{\beta}})$
with its lower bound (see \eqref{eq:eiglb} and \eqref{eq:b}) that is obtained from the signal power spectrum we get
the sufficient condition in Theorem~\ref{thm:approximately-sparse}.

\subsection{Corollary~\ref{cor:full}}
We consider $k = \Theta(p)$. In \eqref{eq:b} the eigenvalues of the autocorrelation
matrix are sorted. They are $n$ smallest eigenvalues out of $p$. To tighten the condition
in Theorem~\ref{thm:sufficient-condition-tight} we use a lower bound in \cite{Merikoski1997}
on the product of $n$ smallest eigenvalues of the matrix instead of lower bounding \eqref{eq:b}
by replacing the eigenvalues with the minimum eigenvalue.
Therefore, we lower bound \eqref{eq:b} as
\begin{equation}
\begin{split}
b &\geq (p - n) \log \left ( \frac{p - n}{\mathrm{SNR}} \right ) +  \log \left | R_{\tilde{\beta}}\right |\\
            &= (p - n) \log \left ( \frac{p - n}{\mathrm{SNR}} \right ) + 
            \left [\frac{n}{2\pi} \int_{0}^{2\pi} \log S(\omega)\; \mathrm{d}\omega \right]
\end{split}
\end{equation}
where the second term in the right-hand side is the integral of
the logarithm of the signal power spectrum \cite{Merikoski1997,Gray2006}.
Noting that inequalities $k \leq p$ and $n \geq k$ hold, $k = \Theta(p)$ implies that 
$$G = \frac{1}{2\pi} \int_{0}^{2\pi} \log S(\omega)\; \mathrm{d}\omega.$$

\section{Conclusion}
\label{sec:conclusion}
In this paper we considered the recovery of the ergodic wide-sense stationary sparse signals.
This is of great importance for computing bounds where the signal is approximately sparse.
In such practical situations finding the infimum of the signal power spectrum and computing
the sufficient condition in Theorem~\ref{thm:approximately-sparse} is straightforward.

We computed a sufficient condition that is at least as tight as the sufficient condition in \cite{Rad2011}
and the necessary condition in \cite{Wang2010}.
Corollary~\ref{cor:sufficient-condition} holds for all signal-to-noise ratios whereas the sufficient condition in \cite{Rad2011}
requires specific signal-to-noise ratio regimes \cite{Rad2011}.
Therefore, the given sufficient condition in this paper improves the previous results in
\cite{Rad2011,Wainwright2009,Tarokh2010,Fletcher2009} and outperforms the necessary condition in \cite{Wang2010}.
\section*{Acknowledgment}
The author would like to thank Prof. Claude Berrou for careful reading of an early version of this manuscript and fruitful
discussions.
\bibliographystyle{IEEEtran}
\bibliography{biblio}
\end{document}